\PassOptionsToPackage{desactivate}{linenoaa}
\documentclass[onecolumn]{aa}
\makeatletter
\def\@floatboxreset{\@floatpenalty\@M\centering}
\makeatother
\usepackage[stable]{footmisc}
\usepackage{amsmath}
\usepackage{amsfonts}
\usepackage{textgreek}
\usepackage{natbib}
\bibpunct{(}{)}{;}{a}{}{,}
\providecommand{\sorthelp}[1]{}

\usepackage[varg]{txfonts}
\usepackage{placeins}
\usepackage{graphicx}
\usepackage{epstopdf}
\usepackage{ifthen}
\usepackage{mathtools}
\usepackage[breaklinks, colorlinks, citecolor=blue]{hyperref}
\usepackage{url}
\usepackage[table,dvipsnames]{xcolor}
\usepackage{physics}
\usepackage{silence}

\usepackage[utf8]{inputenc}
\usepackage[english]{babel}
\definecolor{linkcolor}{rgb}{0.0,0.3,0.5}
\hypersetup{
    unicode,
    colorlinks=true,
    linkcolor=linkcolor,
    citecolor=linkcolor,
    filecolor=linkcolor,
    urlcolor=linkcolor,
}
\usepackage{color,colortbl}
\usepackage{tensind}
\tensordelimiter{?}
\DeclareGraphicsExtensions{.bmp,.png,.jpg,.pdf}
\usepackage{verbatim}
\usepackage[normalem]{ulem}
\usepackage{orcidlink}
\usepackage{soul}
\usepackage{standalone} 
\usepackage{tikz}

\usetikzlibrary{shapes.geometric, arrows.meta, positioning, fit, backgrounds, calc, angles, quotes}
\usepackage{xspace}
\renewcommand{\coloneqq}{\mathrel{\mathop:}=} 
\usepackage{array}
\usepackage{multirow}
\usepackage{makecell}
\AtBeginDocument{}
\definecolor{burntsienna}{rgb}{0.91, 0.45, 0.32}

\newcommand{\quickpol}{\texttt{QuickPol}\xspace}
\newcommand{\smarties}{\texttt{SMARTIES}\xspace}
\newcommand{\ducc}{\texttt{ducc}\xspace}
\newcommand{\lbs}{\texttt{LBS}\xspace}

\newcommand{\sforp}{\texttt{Smarties4Planck}\xspace}

\newcommand{\healpix}{\textsc{HEALPix}\xspace}

\providecommand{\Planck}{\textit{Planck}\xspace}
\providecommand{\planck}{\Planck}

\begin{document}

\title{Exact map-space beam convolution via spin-mapmaking}
\titlerunning{Exact map-space beam convolution via spin-mapmaking}
\author{
Camille Demay\orcidlink{0009-0006-0101-8677}\inst{1,2,3}~\thanks{\email{camille.demay@unife.it}}
\and
Magdy Morshed\orcidlink{0000-0002-3214-8881}\inst{3}~\thanks{\email{magdy.morshed@fe.infn.it}}
\and
Martina Gerbino\orcidlink{0000-0002-3538-1283}\inst{3}~\thanks{\email{gerbinom@fe.infn.it}}
\and
Luca Pagano\orcidlink{0000-0003-1820-5998}\inst{1,3,4}~\thanks{\email{luca.pagano@unife.it}}
}

\institute{
\small Dipartimento di Fisica e Scienze della Terra, Universit\`a degli Studi di Ferrara, via Saragat 1, I-44122 Ferrara, Italy\goodbreak
\and
Dipartimento di Fisica, Università degli studi di Trento, via Sommarive 14, I-38123 Trento, Italy
\and
Istituto Nazionale di Fisica Nucleare, Sezione di Ferrara, via Saragat 1, I-44122 Ferrara, Italy\goodbreak
\and
Institut d'Astrophysique Spatiale, CNRS, Univ. Paris-Sud, Universit\'{e} Paris-Saclay, B\^{a}t. 121, 91405 Orsay cedex, France
}

\date{\vglue -1.5mm \today \vglue -5mm}

\abstract{
Forthcoming cosmic microwave background (CMB) experiments are targeting ambitious science goals, including the exploration of the particle content of the Universe across cosmic times and the search for the faint signal from primordial gravitational waves. To pursue these goals, CMB experiments will house an unprecedented high number of detectors that will dramatically enhance their sensitivity. This will in turn require a tight control of instrumental systematics, demanding efficient and precise simulation of systematic effects. Leveraging spin-mapmaking, we present a map-based method to propagate beam convolution systematics to the map-level, especially well-suited for main-beam studies. From the scanning strategy information compressed into precomputed \textit{h}-maps and the beam and sky harmonic coefficients, our method directly outputs beam convolved maps. We validate against pixel-centered TOD-level $4\pi$ convolution for circular, elliptical, and \planck HFI scanning beams, demonstrating that our method retains agreement to numerical precision, while operating in map space. This formalism enables efficient simulation of beam convolution systematics across the full focal plane of upcoming experiments. As a first practical application, we demonstrate the ability to efficiently produce beam-convolved simulated \planck maps, showing excellent agreement with the \quickpol pipeline at the power-spectrum level. We release a public code, \sforp, to easily generate such maps for broader use.}

\authorrunning{Demay et al.}

\maketitle

\section{Introduction}
\label{sec:intro}
The cosmic microwave background (CMB) anisotropies have proven to be an exceptionally powerful probe of the standard model of cosmology over the past decades \citep[for a review, see e.g.][]{Durrer:2015lza}. Decisive milestones were achieved by satellite missions such as WMAP \citep{hinshaw2012} and \planck \citep{planck2016-l01, planck2016-l06}, as well as ground-based facilities like the Atacama Cosmology Telescope (ACT) \citep{AtacamaCosmologyTelescope:2025blo,AtacamaCosmologyTelescope:2025nti}, the South Pole Telescope (SPT) \citep{SPT-3G:2025bzu}, and the degree-scale focused experiments POLARBEAR/Simons Array \citep{P_A_R_Ade_2014} and BICEP/Keck \citep{BICEP:2021xfz}. These experiments have mapped temperature and polarization anisotropies to cosmic variance limits over a wide range of angular scales.
Building on these successes, the focus of current and next-generation observational efforts has definitively shifted toward two main frontiers. The first is the precise measurement of the large-to-degree-scale CMB polarization signal to detect primordial B-modes, i.e. a distinct signature of the stochastic background of gravitational waves predicted by many inflationary models \citep{Kamionkowski:2015yta}. The second frontier is the exploration of small-scale physics in the temperature and polarization damping tail, where high-resolution observations are crucial for constraining neutrino masses, light relics, and the structural growth of the universe through secondary anisotropies like CMB lensing. To achieve these ambitious scientific goals, upcoming ground-based and space-borne observatories, such as the Simons Observatory \citep{ade_simons_2019}, and LiteBIRD \citep{LiteBIRD:2022cnt}, will deploy large focal planes with $O(10^4 - 10^5)$ detectors and integrate over several years of observation.

Achieving the stringent constraints targeted by these upcoming experiments will require unprecedented control over both astrophysical foregrounds and instrumental systematics. Among the most critical instrumental effects is the precise characterization of detector beams. Complex optical systems inevitably yield non-ideal, asymmetric beam shapes. If not properly accounted for, the convolution of these non-circular beams with the local sky induces severe temperature-to-polarization ($T \rightarrow P$) leakage, which can easily overwhelm the faint primordial B-mode signal and bias small-scale parameter extraction. The standard way to study such systematic effects in CMB experiments is to run Time-Ordered Data (TOD) simulations, computing the timestream of each detector at each time step. For non-circular beams, this rigorously requires computing a $4\pi$ convolution between the beam and the sky at every integration step. Given the massive detector counts and high sampling rates of future projects, exact $4\pi$ TOD simulation becomes a severe computational bottleneck, demanding exorbitant processing power and memory.

Although there is no way to fully replace realistic TOD simulations for all applications -- particularly regarding time-dependent effects such as cosmic ray glitches, thermal fluctuations, or atmospheric noise -- many stationary systematics can be efficiently studied directly at the map level. Early efforts to bypass brute-force TOD generation focused on fast spherical harmonic transforms to speed up timestream generation \citep{Wandelt:2001gp,Challinor:2000df,Prezeau:2010mx,Duivenvoorden:2018zdp}, or on decomposing asymmetric main beams into combinations of symmetric Gaussian kernels \citep[e.g., \textsc{Asymfast};][]{Tristram:2003ee}. In the map domain, pixel-space techniques like \textsc{FEBeCoP} \citep{mitra2010} introduced effective asymmetric beams built pixel by pixel from pointing histories. While these methods marked major advances, pixel-space matrices become memory-prohibitive at high resolutions, while TOD-based harmonic convolutions still require synthesizing vast timestreams. 

More recently, developments in spin-mapmaking and harmonic-domain techniques have shown that asymmetric beams and scan coverage can be captured without explicitly simulating timestreams or storing massive pixel matrices \citep{McCallum:2021oxv,Takase:2025bkc,Morshed_2026}.
Building on this framework, we present in this paper a novel and efficient formalism to simulate beam-convolved temperature and polarization maps directly by using spin-mapmaking. Unlike previous works, which relied on perturbative expansions around an ideal circular beam to study small departures from circularity \citep{Fabbian:2021hlw,Takase:2025bkc,Mirmelstein:2020pfk}, our approach computes the exact convolved maps directly from the spherical harmonic coefficients of both the beam and the sky. We achieve this by coupling their harmonic-space representation to the experimental scanning strategy in real space via precomputed orientation maps, hereafter $h$-maps (defined in Section~\ref{sec:formalism}), which encode the scanning strategy's coverage of orientation angles at each sky pixel, to capture the full angular dependence of the telescope coverage on the sky. This allows for fast exact map-level convolution that scales efficiently for next-generation experiments. The paper is organized as follows. In Section \ref{sec:formalism}, we lay down the mathematical formalism of the $h$-map beam convolution and its relation to spin-mapmaking. In Section \ref{sec:validation}, we validate the accuracy and computational efficiency of our formalism by comparing it against full TOD simulations with exact $4\pi$ convolution. As a first practical application, Section \ref{sec:planck} applies this methodology to \planck data, comparing the angular power spectra derived from our convolved maps against those generated by the established \quickpol pipeline \citep{Hivon:2016qyw}. Finally, we discuss the implications for future surveys and conclude in Section \ref{sec:conclusion}.

\section{Formalism}
\label{sec:formalism}
\subsection{Spin-mapmaking}
CMB experiments aim to extract the temperature and polarization information from the timestream of a detector by leveraging the fact that temperature is a spin-0 signal while polarization is a spin-2 signal. In the absence of polarization modulation, such as the one induced by an half-wave plate (HWP) rotating continuously or in steps, experiments must rely on the scanning strategy and the rotation of the sky relative to the instrument to separate these signals. The noiseless timestream of a polarized detector scanning the sky can be modeled as:
\begin{equation}
     d({\Omega}_t, \zeta_t)=s \left[ {T} + \gamma\left(\frac{{Q} + i{U} }{2}\right) e^{ -2i (\zeta_t +\psi)} + \gamma \left(\frac{{Q} - i{U} }{2}\right) e^{2i (\zeta_t +\psi)} \right] +\text{systematics},
    \label{eq:ideal_tod}
\end{equation}
where ${T}$, ${Q}$, and ${U}$ are the sky Stokes parameters convolved with the instrument's nominal beam profile, sampled along the time-dependent pointing direction $\Omega_t = (\theta_t, \varphi_t)$. Following \cite{Jones:2006ac}, the polarization efficiency $\gamma$ of the detector and gain calibration constant $s$ are explicitly factored out\footnote{In the following treatment, to keep the equations tidy, we assume perfectly calibrated timestreams and unitary polarization efficiency, i.e. $s=\gamma=1$ in Eq.~\ref{eq:ideal_tod}. These effects are straightforward to implement and are accounted for in the companion code of this paper.}. The field spins are defined by their transformation properties under a rotation of the local coordinate system by the parallactic angle $\zeta_t$. Specifically, $\zeta_t$ is defined in the plane tangent to the detector's pointing direction as the angle between geographical North and the reference axis used to define the detector orientation angle $\psi$. Generically, a term of spin $n$ transforms as $f(\Omega, \zeta, \psi) = \bar{f}(\Omega, \psi) e^{-in\zeta}$.
Equivalently to the standard binning approach to map-making \citep[see, e.g.][]{Tegmark:1996qs}, the Stokes parameters can be recovered using a spin-based mapmaking approach \citep{McCallum:2020jsp}. Following the formulation derived in \cite{Morshed_2026} for a focal plane comprising $N_{\text{det}}$ detectors, the spin-based mapmaking equation, which we implement throughout this paper using the Spin MAp Reduction Toolkit for systematics Impact EStimation (\smarties)\footnote{\url{https://github.com/simonsobs/smarties}} package, is given by:

\begin{gather}
    \begin{pmatrix}
    {\hat{T}} \\
    {\hat{Q} - i\hat{U}} \\
    {\hat{Q} + i \hat{U}}
    \end{pmatrix}
    =
    \begin{pmatrix}
    \tilde{h}^{\text{tot}}_{0} & \frac{1}{2} \tilde{h}^{\text{tot}}_2 & \frac{1}{2} \tilde{h}^{\text{tot}}_{-2} \\
    \frac{1}{2} \tilde{h}^{\text{tot}}_2 & \frac{1}{4} \tilde{h}^{\text{tot}}_{4} & \frac{1}{4} \tilde{h}^{\text{tot}}_{0} \\
    \frac{1}{2} \tilde{h}^{\text{tot}}_{-2} & \frac{1}{4} \tilde{h}^{\text{tot}}_{0} & \frac{1}{4} \tilde{h}^{\text{tot}}_{-4}
    \end{pmatrix}^{-1}
    \begin{pmatrix}
    \sum_{j=1}^{N_{\text{det}}} \frac{N^j_{\text{hits}}}{N^{\text{tot}}_{\text{hits}}} \sum_{k' = - \infty}^{\infty} \tilde{h}_{0 - k'}^j \tilde{S}_{k'}^j \\
    \frac{1}{2}\sum_{j=1}^{N_{\text{det}}} e^{i2\psi^j} \frac{N^j_{\text{hits}}}{N^{\text{tot}}_{\text{hits}}} \sum_{k' = - \infty}^{\infty} \tilde{h}_{2 - k'}^j \tilde{S}_{k'}^j \\
    \frac{1}{2}\sum_{j=1}^{N_{\text{det}}} e^{-i2\psi^j} \frac{N^j_{\text{hits}}}{N^{\text{tot}}_{\text{hits}}} \sum_{k' = - \infty}^{\infty} \tilde{h}_{-2 - k'}^j \tilde{S}_{k'}^j
    \end{pmatrix},
    \label{eq:mapmaking}
\end{gather}
where $\hat{T}$, $\hat{Q}$, and $\hat{U}$ are the estimated maps, and $N_{\text{hits}}^{\text{tot}}$ and $N_{\text{hits}}^j$ are the number of data points taken in a pixel, respectively across all detectors and by detector $j$. The individual detector $h$-maps, $\tilde{h}_n^j$, are analogous to the quantities defined in \cite{McCallum:2020jsp, Wallis:2016bja, McCallum:2021oxv}, but with the explicit polarization angle dependence factored into the total focal plane $h$-maps, $\tilde{h}^{\text{tot}}_{n}$. They are defined as
\begin{equation}
\tilde{h}_n^j (\Omega) \coloneqq \frac{1}{N^j_{\text{hits}}}\sum_{i=1}^{N^j_{\text{hits}}} e^{in \zeta_i}
    \qquad \text{and}  \qquad
    \tilde{h}^{\text{tot}}_{n} \coloneqq \sum_{j=1}^{N_{\text{det}}}\frac{N^j_{\text{hits}}}{N^{\text{tot}}_{\text{hits}}} e^{in\psi^j} \tilde{h}_n^j.
    \label{eq:h_maps}
\end{equation}
where the individual detector $h$-maps encode the scan strategy coverage for a given pixel observed with $N_{\text{hits}}^j$ data points. By construction, $\tilde{h}_0^j=1$ for every detector, and hence $\tilde{h}^{\text{tot}}_0=1$.

The magnitude of these $h$-maps serves as a metric for scanning uniformity and cross-linking. A pixel observed uniformly over a wide range of parallactic angles will have highly suppressed higher-order $h$-maps ($n \neq 0$), which results in less mixing between spin components when performing the mapmaking.
The quantities $\tilde{S}_{k'}^j$ represent the spin sky maps, obtained via a discrete Fourier transform of the timestream components, eq.~\eqref{eq:ideal_tod}, with respect to $\zeta$:
\begin{equation}
    d(\Omega_t, \zeta_t) = \sum_{n\in \mathbb{Z}} \tilde{S}_{n}^j (\Omega_t) e^{-in\zeta_t}.
\label{eq:spin-timestream}
\end{equation}
These spin maps encompass both the true underlying sky signals and potential instrumental systematics. In this framework, the discrete convolution between the $h$-maps and the spin maps governs the level of temperature-to-polarization ($T \rightarrow P$) leakage in the final maps estimation. This formalism natively accommodates ``scan-synchronous'' or frame-stationary systematics, i.e. effects that lack intrinsic time dependence and vary solely due to the spatial scanning path. This property allows them to be cleanly cast as additional spin-map components in Eq.~\eqref{eq:ideal_tod}\footnote{Some classes of time-dependent systematics can also be treated by decomposing the timestream into smaller chunks, over which they can be described as stationary, and applying the spin-mapmaking independently to each chunk.}. A prominent class of instrumental systematic satisfying this condition is beam asymmetry. In the following section, we introduce an exact method to compute the effects of non-ideal beam convolutions directly within this $h$-map framework.

\subsection{Beam convolution with spin-mapmaking}
To study systematic effects arising from deviations from an ideal, circularly symmetric Gaussian beam, a common approach is to perform a perturbative expansion around the nominal beam profile. This framework was developed in \cite{ Wallis:2016bja,Mirmelstein:2020pfk} and subsequently applied to map-based simulations in \cite{McCallum:2021oxv} and \cite{Morshed_2026}. Under this perturbative scheme, non-zero beam asymmetry (e.g., slight ellipticity) injects spatial derivative terms of the smoothed temperature and polarization fields directly into the timestream. Although this perturbative approach is highly accurate for small perturbations, which are expected in ground-based experiments \citep[e.g.][for the Simons Observatory]{Gudmundsson:2020wja}, it becomes less reliable for spaceborne missions. The compact, off-axis optical geometries typical of space instruments often yield significantly non-circular beam profiles; for instance, the \planck LFI beams exhibited ellipticities up to 30\% \citep{planck2014-a05}. In the following, we outline an exact method for computing beam-convolved maps using a joint formulation of beam harmonic coefficients and $h$-maps.
In order to incorporate a generalized detector beam response into Eq.~\eqref{eq:ideal_tod}, we decompose both the sky signal and the instrument's beam response into a spherical harmonic basis. Following the derivation in \cite{Hivon:2016qyw}, the noiseless timestream of a monochromatic detector scanning a polarized sky (neglecting circular polarization) is given by\footnote{Eq.~(28) of \cite{Hivon:2016qyw}.}:
\begin{equation}
    d(\Omega_t, \zeta_t) = \sum_{\ell=0}^{\infty} \sum_{m=-\ell}^{\ell} \sum_{n=-\ell}^\ell \left[ {}_{0}a_{\ell m} \,{}_{0}b_{\ell n}^\dagger + \frac{1}{2} \left( {}_{2}a_{\ell m} \,{}_{2}b_{\ell n}^\dagger + {}_{-2}a_{\ell m}\, {}_{-2}b_{\ell n}^\dagger \right) \right] (-1)^n \sqrt{\frac{4\pi}{2\ell +1}} e^{i n \zeta_t} {}_{-n}Y_{\ell m}(\Omega_t),
\label{eq:beam-convolved-tod}
\end{equation}
where ${}_{0,\pm2}a_{\ell m}$ are the standard spin-weighted spherical harmonic coefficients of the sky, and ${}_{n}Y_{\ell m}$ represent the spin-weighted spherical harmonics. Conversely, ${}_{0,\pm2}b_{\ell n}$ are the spin-weighted harmonic coefficients characterizing the detector beam response, which encapsulate both the intrinsic beam geometry and the orientation of the detector's polarization sensitive axis. Commuting the sum indices, this timestream can be expressed as:
\begin{equation}
    d(\Omega_t, \zeta_t) = \sum_{n \in \mathbb{Z}} \left[ \sum_{\ell=|n|}^{\infty} \sum_{m=-\ell}^{\ell} (-1)^{n} \sqrt{\frac{4\pi}{2\ell +1}} \left( {}_{0}a_{\ell m} \, {}_{0}b_{\ell, -n}^{\dagger} +\frac{1}{2}  {}_{2}a_{\ell m} \,{}_{2}b_{\ell, -n}^\dagger +\frac{1}{2} {}_{-2}a_{\ell m} \,{}_{-2}b_{\ell, -n}^\dagger \right)\, {}_{n}Y_{\ell m}(\Omega_t)\right] e^{-in \zeta_t}.
\end{equation}
By comparing this expansion to Eq.~\eqref{eq:spin-timestream}, we can identify the effective spin maps required for the mapmaking equation, Eq.~\eqref{eq:mapmaking},
\begin{equation}
    \tilde{S}_{n} (\Omega)= \sum_{\ell=|n|}^{\infty} \sum_{m=-\ell}^{\ell} (-1)^{n} \sqrt{\frac{4\pi}{2\ell +1}} \left( {}_{0}a_{\ell m} \, {}_{0}b_{\ell, -n}^{\dagger} +\frac{1}{2}  {}_{2}a_{\ell m} \,{}_{2}b_{\ell ,-n}^\dagger +\frac{1}{2} {}_{-2}a_{\ell m} \,{}_{-2}b_{\ell, -n}^\dagger \right)\, {}_{n}Y_{\ell, m}(\Omega).\label{eq:expr_spin_maps}
\end{equation}
We give an interpretation of these spin maps in terms of systematics terms in Appendix~\ref{sec:syst_terms_expansion}. With this, convolution of spin maps with $h$-maps via Eq.~\eqref{eq:mapmaking} provides an exact, map-based, non-perturbative method to evaluate reconstructed Stokes parameters for arbitrary beam profiles. In practice, we compute these spin maps from the harmonic coefficients through spin-weighted spherical harmonic transforms.

Existing $4\pi$ convolvers evaluate Eq.~\eqref{eq:beam-convolved-tod} at each time step, relying on highly optimized computations of Wigner matrices to describe the rotations of the detector beams during the scan \citep{Prezeau:2010mx,reinecke2006,Wandelt:2001gp}. Our approach is closer to that of \cite{Duivenvoorden:2018zdp}, which computes spin maps similar to those in Eq.~\eqref{eq:expr_spin_maps} and then applies an $e^{-in \zeta_t}$ pointing factor to obtain the TOD of Eq.~\eqref{eq:beam-convolved-tod}. The novelty of our approach lies in precomputing this pointing information into the $h$-maps, making the method fully map-based and independent of TOD length. While the derivation presented here assumes scanning without a half-wave plate, it can easily be extended to include one by following the approach of \cite{Takase:2025bkc}.

The range of $h$-maps required by the discrete convolution of Eq.~\eqref{eq:mapmaking} extends up to $|n|=m_{max}+2$, two spin orders beyond the beam's own azimuthal cutoff $m_{max}$, due to the $\pm2$ spin shifts coupling the spin maps to the output Stokes parameters. Thus, each detector requires storing $m_{max}+2$ complex-valued maps for a beam with no particular symmetry, or half that if the beam is confined to even azimuthal orders $m$ thanks to an axis of symmetry, as is the case for elliptical (and circular) Gaussian beams. Keeping $m_{max}$, and hence this storage cost, small is possible either because the beam harmonic coefficients decay rapidly with $|m|$, as is naturally the case for main-beam convolution, or because the scanning strategy itself suppresses the higher-order $h$-maps, as explored by \cite{Thomas:2021gef} for ground-based experiments, and by \cite{LiteBIRD:2024fzn} for LiteBIRD. We return to this trade-off quantitatively in Appendix~\ref{sec:mmax_discussion}. 
Since $h$-maps are stored independently per detector, the total cost scales with $N_{\text{det}}$ and can become substantial for the large focal planes and high $n_{side}$ typical of next-generation experiments, even when $m_{max}$ remains modest: for a \healpix\footnote{\url{http://healpix.sourceforge.net}.} pixelization at $n_{side}=2048$, the storage cost of a single $h$-map for one detector and one spin is approximately 768 MB in double precision. In this work, we focus on validation (Section~\ref{sec:validation}) and application to \planck (Section~\ref{sec:planck}), which involve only a few tens of detectors, so storage is not a limiting factor. However, reducing the storage footprint will be relevant for full-focal-plane simulations of current and next-generation experiments, especially at higher $m_{max}$. To this end, \cite{Morshed_2026} explored, in the context of the Simons Observatory, the possibility of saving $h$-maps at low resolution and/or only for a few detectors, and then interpolating them on the fly.

\section{Validation with respect to \texorpdfstring{$4\pi$}{4pi} convolution}\label{sec:validation}
In this section, we validate this map-based approach to beam convolution against the traditional $4\pi$ convolution of the timestream \citep{Wandelt:2001gp,Prezeau:2010mx}. To do so, we compare the output maps of the $h$-map based convolution to a binned map obtained from a $4\pi$ convolved timestream. We test the approach for three classes of beams of increasingly complex profile: circular Gaussian (for a validation in the ideal case), elliptical Gaussian (for a semi-realistic case in presence of small departures from the circular profile), and \planck HFI scanning beams \citep{planck2014-a08} (for a real-case scenario with large departures from ideality).

\subsection{TOD and map-based simulation setups}
\label{sec:setup}

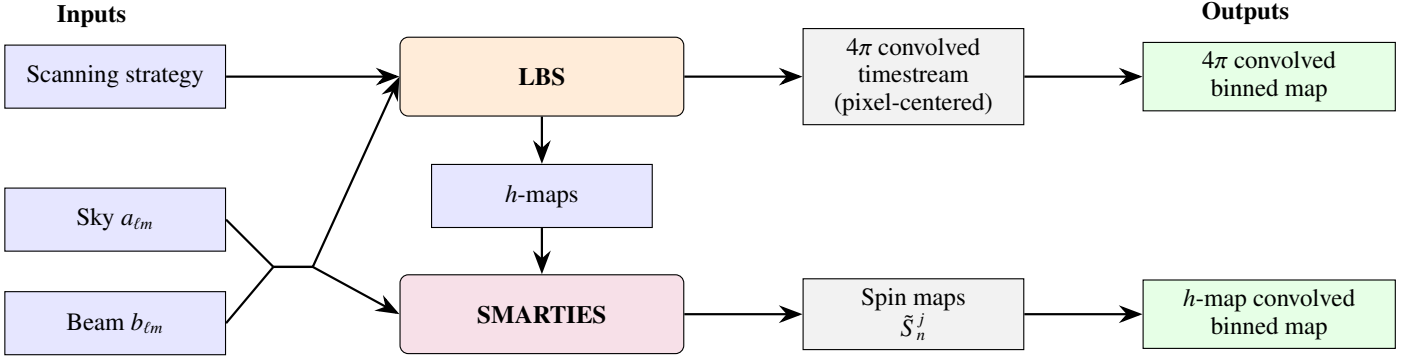
\begin{figure}[!htbp]
\begin{tikzpicture}[
    node distance=0.8cm and 1.2cm,
    % Block styles
    input/.style={rectangle, draw=black, fill=blue!10, minimum width=2.8cm, minimum height=0.8cm, align=center, font=\small},
    process/.style={rectangle, draw=black, fill=orange!15, minimum width=3.2cm, minimum height=0.9cm, align=center, font=\small\bfseries, rounded corners=3pt},
    output/.style={rectangle, draw=black, fill=green!10, minimum width=3.2cm, minimum height=0.8cm, align=center, font=\small},
    product/.style={rectangle, draw=black, fill=gray!10, minimum width=2.8cm, minimum height=0.8cm, align=center, font=\small},
    arrow/.style={-{Stealth[length=3mm]}, thick},
    dasharrow/.style={-{Stealth[length=3mm]}, thick, dashed},
    headlessarrow/.style={-, thick},
    % Box styles
    box/.style={draw=black, dashed, rounded corners=8pt, inner sep=10pt},
    label/.style={font=\small\bfseries, text=black},
]
    
% ===== INPUTS =====
\node[input] (scan) {Scanning strategy};
\node[input,below=1.0cm of scan] (alms) {Sky $a_{\ell m}$};
\node[input, below=0.5cm of alms] (blms) {Beam $b_{\ell m}$};

\coordinate[right=0.6cm of alms.east,yshift=-0.6cm] (junction_skybeam);

% ===== LITEBIRD_SIM BOX =====
\node[process, right=2.2cm of scan, minimum height=1cm, minimum width=3.6cm] (lbs_main) {LBS};

% ===== SMARTIES BOX (below LBS) =====
\node[process, below=2cm of lbs_main, minimum height=1cm, minimum width=3.6cm, yshift=0cm, fill=purple!12.] (smarties_main) {SMARTIES};

% ===== LBS PRODUCTS =====
\node[product, right=1.5cm of lbs_main, yshift=0.0cm] (timestream) {4$\pi$ convolved\\timestream \\ (pixel-centered)};
\node[input, below=0.6cm of lbs_main, yshift=0.0cm] (hmaps) {$h$-maps};

% ===== SMARTIES OUTPUT =====
\node[product, right=1.5cm of smarties_main, xshift=0.0cm] (spinmaps) {Spin maps\\$\tilde{S}^j_n$};

% ===== FINAL PRODUCTS =====
\node[output, right=1.5cm of timestream, xshift=0.0cm, yshift=0.0cm] (refmap) {4$\pi$ convolved\\binned map};
\node[output, right=1.5cm of spinmaps, xshift=0.0cm] (hconvmap) {$h$-map convolved\\binned map};
 
% ===== ARROWS: Inputs to LBS =====
\draw[headlessarrow] (alms.east) -- (junction_skybeam);
\draw[headlessarrow] (blms.east) -- (junction_skybeam);
\draw[arrow] (scan.east) --  node[below, pos=0.3, font=\footnotesize] {} (lbs_main.west);

% ===== ARROWS: LBS products =====
\draw[arrow] (lbs_main.east) --  (timestream.west);
\draw[arrow] (lbs_main.south) -- (hmaps.north);

% ===== ARROWS: LBS to reference map =====
\draw[arrow] (timestream.east) --  (refmap.west);

% ===== ARROWS: h-maps to Smarties =====
\draw[arrow] (hmaps.south) --  (smarties_main.north)  ;

% ===== ARROWS: junction to LBS =====
\draw[arrow] (junction_skybeam) -- ++(0.5,0) -- (lbs_main.west);

% ===== ARROWS: junction to Smarties =====
\draw[arrow] (junction_skybeam) -- ++(0.5,0) -- (smarties_main.west);

% ===== ARROW: Smarties to spin maps =====
\draw[arrow] (smarties_main.east) -- (spinmaps.west);

% ===== ARROWS: spin maps + h-maps to final mapmaking =====
\draw[arrow] (spinmaps.east) -- ++(0.5,0) |- (hconvmap.west);

% ===== LABELS =====
\node[label, above=0.1cm of scan, xshift=-0.3cm] {\textbf{Inputs}};
\node[label, above=0.1cm of refmap, xshift=-0.3cm] {\textbf{Outputs}};
\end{tikzpicture}
\caption{Flowchart of the beam convolution pipelines compared in this work using the LiteBIRD Simulation Framework (\lbs) and the map-based simulation framework \smarties. The telescope scanning strategy is processed by \lbs to generate intermediate $h$-maps. These $h$-maps, alongside the shared input sky ($a_{\ell m}$) and beam ($b_{\ell m}$) harmonics, are subsequently utilized by \smarties to produce the final convolved maps. The outputs of the TOD-based, $4\pi$ convolved \lbs pipeline and of the h- and spin-map based pipeline are expected to match up to numerical precision.}
\label{sch:flowchart}
\end{figure}

In order to verify our approach, we run TOD simulations with exact $4\pi$ convolution. This involves looping over the pointing information of the telescope and computing the convolution between the beam and the sky at each time step to obtain the TOD. We use the LiteBIRD Simulation Framework \citep{LiteBIRD:2025tua, LiteBIRD:2025vxv}\footnote{\url{https://github.com/litebird/litebird_sim}.} (\lbs) to run these simulations; therefore, throughout this validation, we adopt the LiteBIRD scanning strategy presented in \cite{LiteBIRD:2026gdf}, with no polarization modulator and a spin rate of 0.3 rpm \cite{litebirdcollaboration2026litebirdmissionoverviewmission}. From the detector beam $b_{\ell m}$, the sky $a_{\ell m}$, and the scanning strategy, \lbs computes the convolved TODs via the optimized convolution tools included in the \ducc\footnote{\url{https://gitlab.mpcdf.mpg.de/mtr/ducc}.} collection. Then it applies a binned mapmaking procedure to the TODs assuming \healpix pixelization\footnote{Some choice of sky pixelization is unavoidable whenever real-space quantities are stored on a discrete grid. We adopt \healpix throughout this paper, because it is the pixelization natively used by \lbs, to bin the TOD and compute the $h$-maps. The spin-mapmaking formalism of Sect.~\ref{sec:formalism}, however, makes no reference to \healpix or to any specific pixelization scheme: it applies unchanged to any grid on which $h$-maps and beam-convolved maps can be self-consistently defined, such as the Cylindrical Equal-Area (CAR) grids used by ground-based pipelines and relevant for experiments such as the Simons Observatory, e.g. \texttt{pixell} (\url{https://github.com/simonsobs/pixell}). In particular, as \smarties was developed within the Simons Observatory collaboration, it supports \texttt{pixell}.}. \smarties performs the $h$-map mapmaking using $h$-maps generated by \lbs. Unlike the TOD simulations, the \textit{h}-map approach is intrinsically insensitive to sub-pixel effects, i.e., to the fact that the observations (``\textit{hits}'') within a pixel will be unevenly distributed \citep[see, e.g., Appendix F of][]{planck2013-p03c}. Thus, in order to make a proper comparison, the pointing directions of each sample in the TOD simulation $\Omega_t$ are aligned with the center of the nearest \healpix pixel. Since an identical set of $b_{\ell m}$ coefficients is provided for both the $h$-map approach and the $4\pi$ convolver, we expect the resulting maps to match up to numerical precision. A schematic flowchart of the convolution pipelines is sketched in Fig.~\ref{sch:flowchart}.

For the validation tests, we consider two pairs of polarized detectors with polarization angles $0^\circ$, $90^\circ$, $45^\circ$, and $135^\circ$. We use a CMB-only sky generated assuming a \planck 2018 fiducial cosmology \citep{planck2016-l06}, with full-sky observations.

\subsection{Beams used in the validation}
We validate our approach with the $b_{\ell m}$ coefficients of three classes of beams: circular Gaussian, elliptical Gaussian, and \planck HFI scanning beams. For a circular Gaussian beam with variance $\sigma^2$ and polarization angle $\psi$, the $b_{\ell m}$ coefficients are \citep{Challinor:2000df}:
\begin{eqnarray}
b_{\ell m}
&=& \sqrt{\frac{2\ell+1}{4\pi}}\,
e^{-\frac{1}{2}\ell(\ell+1)\sigma^2}\,
\delta_{m,0} \\
{}_{\pm 2}b_{\ell,m}
&=& b_{\ell,m\pm2}\,e^{2\sigma^2}e^{\pm i2\psi}.
\end{eqnarray} 
where $\delta$ denotes the Kronecker delta. Consequently, solely $\tilde{S}_{0}$ and $\tilde{S}_{\pm2}$ are non-zero, meaning that only $\tilde{h}_{0}$, $\tilde{h}_{2}$, and $\tilde{h}_{4}$ are required to compute the $h$-map discrete convolution in the vector on the RHS of Eq.~\eqref{eq:mapmaking}.

In the elliptical Gaussian case, the beam requires two additional parameters: the ellipse angle $\varepsilon$ and the ellipticity $e$ defined as the ratio between the major axis, $\sigma_{maj}$, and the minor axis, $\sigma_{min}$. The beam width $\sigma$ is then defined as the geometric mean of the width of the two axes. The $b_{\ell m}$ coefficients are null for odd values of $m$, while for positive even values of $m$ \citep{Ng_2005}\footnote{Compared to \cite{Ng_2005}, we drop the $\ell(\ell+1) \approx \ell^2$ approximation and include the $e^{2\sigma^2}$ polarization factor of \cite{Challinor:2000df}.}:

\begin{eqnarray}
    _0 b_{\ell m}&=&\sqrt{\frac{2\ell+1}{4\pi}}\exp\left(\alpha_\ell\right) I_{m/2} \left ( \beta_\ell\right) e^{-im\varepsilon} \\
    _2 b_{\ell m}&=&\sqrt{\frac{2\ell+1}{4\pi}}\exp\left(\alpha_\ell+2\sigma^2\right)  \left [
    e^{i2(\psi -\varepsilon)}I_{(m+2)/2} \left ( \beta_\ell\right)\right]e^{-im\varepsilon} \\
    _{-2} b_{\ell m}&=&\sqrt{\frac{2\ell+1}{4\pi}}\exp\left(\alpha_\ell+2\sigma^2\right)  \left [ e^{-i2(\psi -\varepsilon)}I_{(m-2)/2} \left ( \beta_\ell\right)\right]e^{-im\varepsilon},
\end{eqnarray}

with  $\alpha_\ell=-\frac{1}{4} \ell \left(\ell+1\right)\left( 1+ \frac{1}{e^2}\right) e\sigma^2$, $\beta_\ell=\frac 1 4  \ell \left(\ell+1\right) \left( 1 - \frac{1}{e^2}\right)e\sigma^2$ and $I_{n}$  the  modified Bessel function of the first kind. \\ 
While in principle all values of $m$ should be retained to fully characterize the beam, in practice 
the $b_{\ell m}$ amplitudes drop sharply with $|m|$, which allows us to cut the expansion off at a small $m_{max}$. Then, in order to compute the discrete convolution of Eq.~\eqref{eq:mapmaking}, we need all the $\tilde{h}_{2k}$ such that $2k \leq m_{max}+2$.
In Appendix~\ref{sec:pert_ell}, we also consider the possibility to approximate the elliptical beam shape with a perturbative Taylor expansion, showing that it yields good results in the limit of small ellipticity.
To validate our approach in a more generic case we consider \planck HFI 100GHz scanning beams \citep{planck2014-a08}. These contain only the temperature response; we assume narrow copolar beams to get the polarization components from the temperature one. \cite{Hivon:2016qyw} shows that in this case:
\begin{equation}
    _{\pm 2} b_{\ell m }= \ _{0} b_{\ell, m \pm 2}e^{\pm 2i \psi}.
\end{equation}
The settings used in our validation are reported in Tab.~\ref{table:params}. We note that our validation used only copolar beams for the sake of simplicity. However, the formalism itself does not assume copolar beams. As discussed in Appendix~\ref{sec:crosspolar}, applying it to beams with a crosspolar component is simply a matter of using the corresponding harmonic coefficients.

\begin{table}
\centering
        \begin{tabular}{ c c c c c c c c}
            Case & Beam & FWHM (arcmin) & $n_{side}$& $\ell_{max}$ & $m_{max}$ & Sampling rate (Hz) & Duration (Days)\\
            \hline \hline
           Circular & Circular & 100 & 128& 256 & - & 1 & 365 \\
            Elliptic 1 & Elliptic, $e=1.1$  & 100 & 128& 256 & 4 & 1 & 365 \\
            Elliptic 2 & Elliptic, $e=1.2$  & 100 & 128 & 256 & 4& 1 & 365 \\
            Elliptic 3 & Elliptic, $e=1.3$  & 100 & 128 & 256&4& 1 & 365 \\
            HFI 1& HFI 100GHz & $\sim9.5$ & 1024 & 2048&4& 30 & 365\\
            HFI 2& HFI 100GHz & $\sim9.5$ & 1024 & 2048&6& 30 & 365\\
            \hline \hline
        \end{tabular}
        \caption{Simulation and instrumental parameters utilized for the validation. We consider three beam configurations: ideal circular Gaussian, elliptical Gaussian with varying ellipticities $e$, and the realistic \planck HFI 100~GHz scanning beam. The parameter $m_{\text{max}}$ denotes the truncation order of the beam harmonic expansion coefficients $b_{\ell m}$ fed to \smarties and \lbs. The map-level comparison for these runs is summarized in Fig.~\ref{fig:summary}. Since HFI beams have a $9.5$~arcmin FWHM \citep{planck2014-a08}, we choose $n_{side}=1024$ so that the beam FWHM is larger than the pixel size\footnote{Note that natively \planck HFI, for consistency across frequency, produces maps at $n_{side}=2048$.}.}
        \label{table:params}
\end{table}

\subsection{Results}
\label{sec:results}
\begin{figure}[h]
    \centering
    \includegraphics[width=1\linewidth]{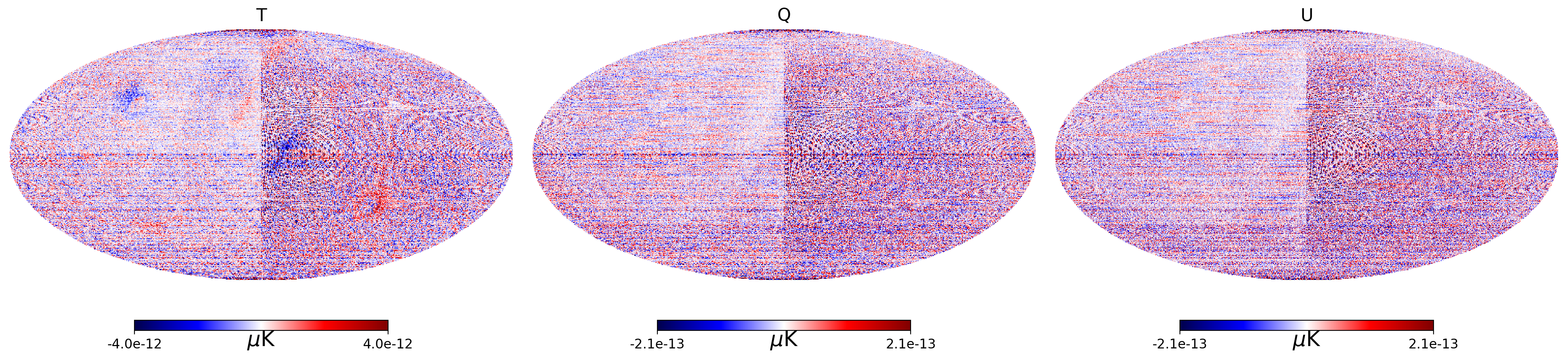}
    \caption{Difference between T,Q,U maps obtained from \lbs $4\pi$ convolution and \smarties with map-based convolution. The maps were generated using the configuration HFI 2 of Tab.~\ref{table:params}.}
    \label{fig:residual_maps}
\end{figure}

\begin{figure}[h]
    \centering
    \includegraphics[width=0.7 \linewidth]{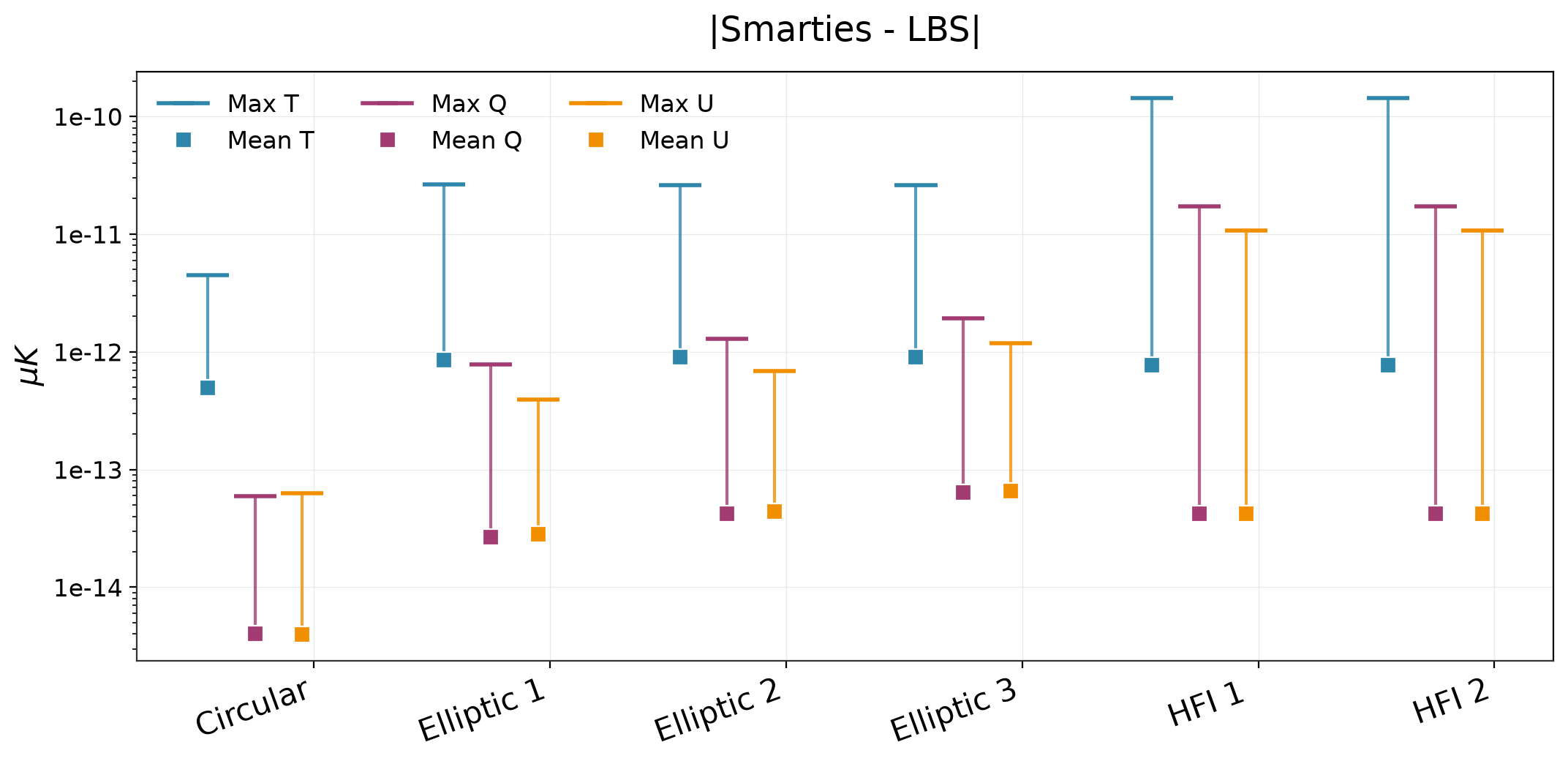}
    \caption{Summary of the agreement between \lbs $4\pi$ convolution and \smarties output maps. We display the mean and maximum of the absolute difference between the T, Q, U maps. The parameters corresponding to each of the configurations on the \textit{x}-axis used to produce the maps are given in Tab.~\ref{table:params}.}
    \label{fig:summary}
\end{figure}

The results are shown in Fig.~\ref{fig:residual_maps}, where we display the residuals between the \lbs and \smarties output maps. In Fig.~\ref{fig:summary}, we summarize the results by highlighting, for each configuration studied, the mean and the maximum of the absolute difference between the T, Q, U maps obtained from the two convolution pipelines. In each case, from the simple circular Gaussian to the observed \planck HFI scanning beams, we recover the same maps with the $4\pi$ convolution using \lbs and the map-based convolution using \smarties, up to numerical precision. As expected, the closest match is in the ideal case of perfectly circular beam: it has a low resolution and the smallest number of spin maps involved, which minimizes the numerical error in the computation of Eq.~\eqref{eq:mapmaking}. Nevertheless, residuals in the elliptical and \Planck HFI beam configurations are always negligible compared to the signal of interest, at the $\lesssim10^{-10}\,\mathrm{\mu K}$ level in T and an order of magnitude smaller in polarization. 

We note that the elliptical and HFI beam validations were performed with a truncated beam harmonic expansion ($m_{max}=4$ or $6$), chosen to keep the number of stored $h$-maps small. In Appendix~\ref{sec:mmax_discussion}, we quantify the impact of this truncation on the reconstructed maps and power spectra, showing that $m_{max}=4$ or $6$ is sufficient for the beam shapes considered in this validation.

This demonstrates that, when provided with the same beam and sky harmonic coefficients ($b_{\ell m}$ and $a_{\ell m}$) and neglecting sub-pixel effects, our approach reproduces the results of a $4\pi$ convolution at a significantly lower computational cost. While the runtime of TOD-level $4\pi$ convolution scales with the number of simulated samples, our map-based approach scales with the number of pixels and the number of spins considered. To put this in perspective, consider an experiment taking three years of data at a 100 Hz sampling rate, producing \healpix maps at $n_{side}=2048$: a single detector timestream would contain $\sim 10^{10}$ samples, while a map contains only $\sim 5 \times 10^7$ pixels. Our approach therefore processes orders of magnitude less data. While the exact speedup depends on the specific implementation, this scaling argument demonstrates that the map-based method can be orders of magnitude faster than TOD-level $4\pi$ convolution.


\section{Application to \planck}
\label{sec:planck}
As a first practical application of our $h$-map spin-mapmaking formalism, we demonstrate its capability to produce synthetic beam-convolved maps for the \planck mission. We benchmark our output against the \quickpol pipeline \citep{Hivon:2016qyw}, which was used extensively in the official \planck data processing and likelihood analyses \citep{planck2016-l06}. \quickpol ingests detector beam harmonic coefficients $b_{\ell m}$ and spatial $h$-maps as input to evaluate the beam properties of the instrument. It operates in harmonic space at the power spectrum level and the output is given directly in terms of spectra. This is a major difference with our formalism that operates at the map level, therefore retaining and propagating more information on the fields of interest, with minimal increase in computational cost. Specifically, \quickpol models the mode mixing between the true input sky power spectra $C_\ell^{X'Y'}$ and the observed pseudo-power spectra $\tilde{C}_\ell^{XY}$ via an effective mixing matrix $W_{\ell}^{XY,X'Y'}$:
\begin{equation}
\tilde{C}_{\ell}^{XY} = \sum_{X'Y'} W_{\ell}^{XY,X'Y'}\, C_{\ell}^{X'Y'},
\label{eq:QP}
\end{equation}

where $X,Y \in \{ T,E,B\}$. The underlying formalism is similar to spin-mapmaking, but in order to propagate directly to the power spectrum level, \quickpol assumes that the $h$-map harmonic coefficients are dominated by low multipoles. While the \quickpol formalism can in principle include scale mixing (see Eq.~38 of \citealt{Hivon:2016qyw}), the implementation based on Eq.~\eqref{eq:QP}, which is used in the \planck analyses and adopted here, does not\footnote{The \quickpol implementation used here is a lightweight wrapper to the \Planck NPIPE QuickPol pipeline: \url{https://github.com/paganol/qp_planck}}. This simplification is valid for smooth scanning strategies such as those of \planck and LiteBIRD and enables very efficient computations.

Using the $h$-maps and beam multipoles $b_{\ell m}$ derived for the eight $100\,\mathrm{GHz}$ detectors of \planck HFI\footnote{These are part of the \Planck NPIPE products available on NERSC at \texttt{/global/cfs/cdirs/cmb/data/planck2020/npipe/aux}.}, we compare the map-level convolution output of \smarties to the spectral predictions of \quickpol.

\begin{figure}
        \centering
    	\includegraphics[width=0.9\linewidth]{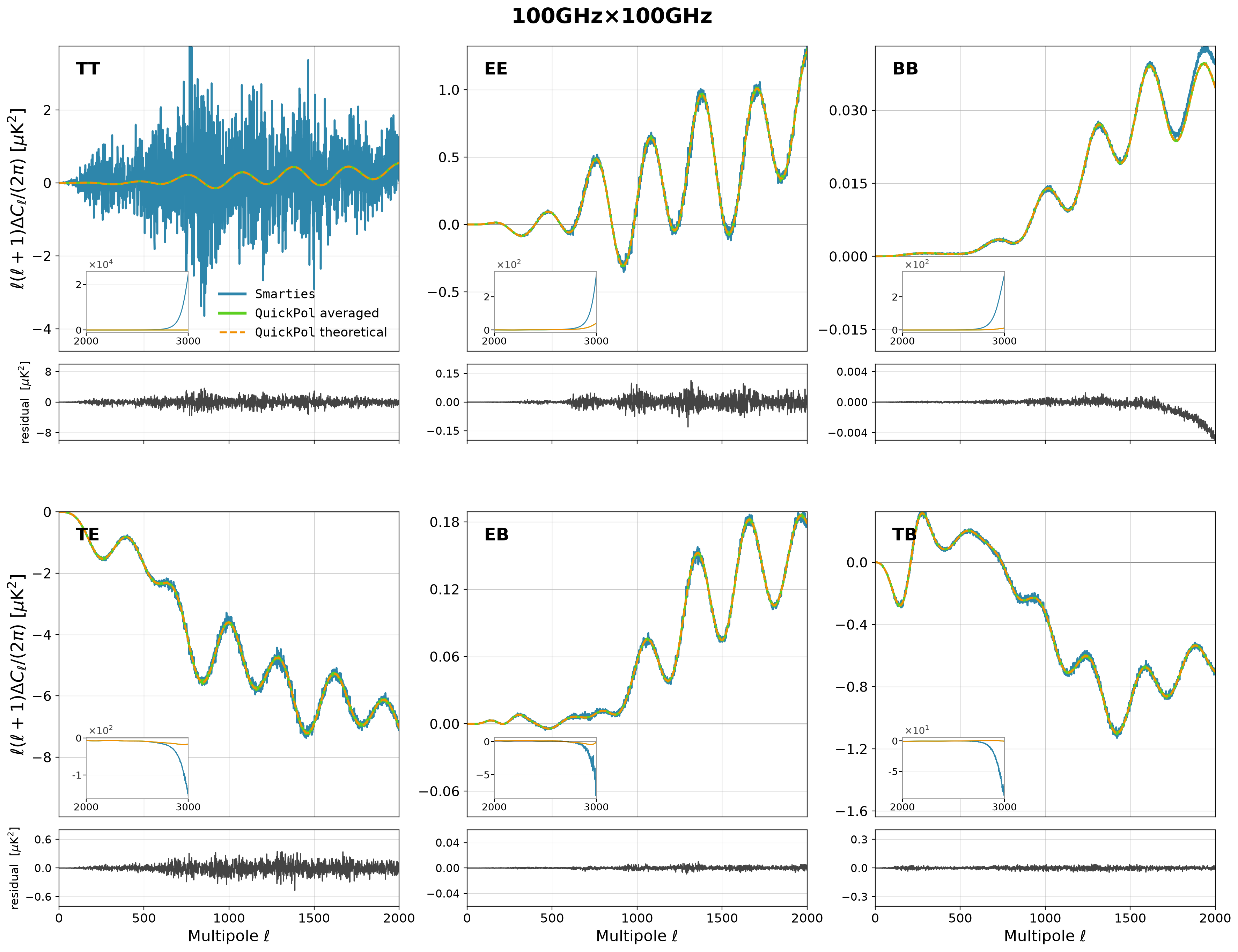}
    	\caption{Bias $\Delta \hat{C}_{\ell}^{XY}$ between the measured and true power spectra for \quickpol and \smarties, using \planck 100 GHz detector $h$-maps and beams. Three cases are shown: \smarties (blue), averaged over 20 CMB realizations with the averaged input spectra as reference; \quickpol averaged over the same realizations (green); and \quickpol run on the theoretical spectrum (orange). Below each panel, we show the residual between the \quickpol theoretical curve and the \smarties curve. The insets show the divergence between the curves at $\ell > 2000$. }
        \label{fig:QPvsSmarties}
    \end{figure}

Figure~\ref{fig:QPvsSmarties} presents the residual bias in the recovered power spectra, defined as
\begin{equation}
    \Delta \hat{C}_{\ell}^{XY} \coloneqq \frac{\hat{C}_{\ell}^{XY}}{W_{\ell}^{XY,XY}} - C_{\ell}^{XY}.
\end{equation}
Three cases are shown, all based on the \planck 2018 best-fit cosmology \citep{planck2016-l06}. For the \smarties curve (blue), $\hat{C}_{\ell}^{XY}$ is extracted from the reconstructed maps using the \textsc{anafast} routine of \healpix and averaged over 20 independent CMB realizations, while $C_{\ell}^{XY}$ is the average of the input spectra computed over the same realizations. For the \quickpol averaged curve (green), both $\hat{C}_{\ell}^{XY}$ and $C_{\ell}^{XY}$ are computed from the averaged input spectra of the same 20 realizations. For the \quickpol theoretical curve (orange), both $\hat{C}_{\ell}^{XY}$ and $C_{\ell}^{XY}$ are computed directly from the theoretical CMB power spectrum. Below each panel, we show the residual between the \quickpol theoretical curve and the \smarties curve. As in the previous sections, we consider a CMB-only sky with full-sky observations.

\medskip
Overall, \quickpol and \smarties show excellent agreement across low and intermediate multipoles. Since \quickpol was validated against full TOD-level simulations and deemed accurate enough for the official \planck analysis, our agreement with \quickpol confirms that our map-based approach is suitable for \planck applications. 
We note that the \smarties curves show noticeably more scatter than the
\quickpol averaged ones. This is a consequence of the different levels
at which the two formalisms operate: \quickpol computes a single mixing matrix
$W_{\ell}^{XY,X'Y'}$ from the sky-averaged $h$-maps, whereas in \smarties the
mapmaking matrix in Eq.~\eqref{eq:mapmaking} is inverted pixel by pixel, so
the effective mixing kernel depends on the local value of the $h$-maps and
couples to the CMB realization through the spin maps $\tilde S_n(\Omega)$. This additional coupling introduces scatter in the computed spectra beyond what is obtained when the \quickpol mixing matrix is applied to a single realization of the CMB.

At high multipoles ($\ell \gtrsim 1500$), a noticeable divergence appears in the $BB$ spectrum. We also observe high-$\ell$ departures in the other spectra (see insets), with \quickpol and \smarties having fully diverged by $\ell=3000$. This high-$\ell$ discrepancy originates from \quickpol's no-scale-mixing approximation, which ignores power leakage from lower multipoles into higher ones. At high $\ell$, where beam window functions heavily suppress the primary sky signal, even small absolute leakage contributions from unattenuated low-$\ell$ modes become proportionally significant. Conversely, complementary tests with elliptical beams confirm a reverse leakage from high-$\ell$ into low-$\ell$ modes, further reinforcing this scale-mixing interpretation (see Appendix~\ref{sec:scale_leakage}). We note, however, that the 100 GHz detectors are not used above $\ell \sim 1200$ in \planck analyses (\citealt{planck2016-l05}, Table~13), so this high-$\ell$ discrepancy has no practical impact on \planck results.

\medskip

We provide a publicly available code, \sforp\footnote{\url{https://github.com/camilledemay/smarties4planck}}, to easily produce convolved \planck maps with this formalism (see Appendix~\ref{sec:planck_code} for details). This tool can be used to generate \planck signal simulations, including CMB and/or foregrounds, with minimal computational effort.

\section{Conclusion}\label{sec:conclusion}
In this work, we presented an efficient, map-based formalism to simulate beam-convolved CMB maps directly in map space, bypassing the computationally expensive $4\pi$ convolution required at the time-ordered data (TOD) level. By formulating beam convolution within the spin-mapmaking framework, we established that a beam-convolved timestream can be modeled as a scan of effective spin sky maps coupled to the instrument scanning strategy via precomputed $h$-maps. The $h$-maps encode the scanning strategy's coverage of orientation angles at each sky pixel. The key theoretical result is the general, non-perturbative expression for these beam-convolved spin sky maps (Eq.~\eqref{eq:expr_spin_maps}).\\
The map-based pipeline is implemented via the \smarties package. We validated it against exact, pixel-centered, $4\pi$ TOD convolutions implemented using the LiteBIRD Simulation Framework \lbs. We demonstrated agreement down to numerical precision across increasingly complex beam shapes: circular Gaussian, highly elliptical Gaussian ($e = 1.3$), and realistic \planck HFI scanning beams. 

As a first practical application, we demonstrated the capability of our framework to efficiently synthesize beam-convolved \planck maps. The derived angular power spectra show excellent agreement with the \quickpol pipeline across low and intermediate multipoles, while accurately preserving high-$\ell$ scale-mixing effects that analytical beam matrix approximations like that assumed in \quickpol omit. Our released open-source software pipeline (\sforp) allows the community to rapidly generate realistic beam-convolved \planck\ signal maps at a fraction of traditional computational cost.

The technique presented in this work provides a scalable, numerically exact alternative to TOD-level $4\pi$ convolutions, trading the one-off precomputation and storage of static $h$-maps for orders-of-magnitude less data to process thanks to the computation being map-based. As mentioned in Section~\ref{sec:results}, a single detector timestream for a next-generation experiment contains $\sim 10^{10}$ samples, while the corresponding $n_{side}=2048$ map contains only $\sim 5 \times 10^7$ pixels. By retaining map-level consistency with respect to TOD level $4\pi$ convolution down to numerical precision, this framework enables end-to-end frame-stationary systematic error propagation across large focal planes for upcoming missions like the Simons Observatory and LiteBIRD, addressing one of the core challenges of new-generation CMB analysis.

\section*{Acknowledgments}
We thank Giulio Fabbian and Adri Duivenvoorden for very useful discussions. Some of the results in this paper have been derived using the \textsc{healpy} \citep{Zonca2019} and \healpix\citep{gorski2005} packages. We acknowledge the use of \texttt{numpy} \citep{harris2020array} and \texttt{matplotlib} \citep{Hunter:2007} software packages. This paper and related research have been conducted during and with the support of the Italian National PhD Programme in Space Science and Technology.
MM, MG and LP are funded by the European Union (ERC, RELiCS, project number 101116027). Views and opinions expressed are however those of the author(s) only and do not necessarily reflect those of the European Union or the European Research Council Executive Agency. Neither the European Union nor the granting authority can be held responsible for them. This work has also received funding by the European Union’s Horizon 2020 research and innovation  program under grant agreement no. 101007633 CMB-Inflate. The authors acknowledge the financial support from the COSMOS network (www.cosmosnet.it) through the ASI (Italian Space Agency) Grants 2016-24-H.0 and 2016-24-H.1-2018. We acknowledge CINECA for the availability of high performance computing resources and support through the CINECA-INFN agreement. The authors acknowledge the use of Anthropic’s Claude AI, Google’s Gemini 3.6 and Deepseek V4.1 as supplementary research-assistance tools for plotting, data analysis, and text editing. 

\bibliographystyle{aa}
\bibliography{bibliography,Planck_bib}

\begin{appendix}
\section{Systematics terms interpretation}
\label{sec:syst_terms_expansion}
In order to build intuition for what the spin maps represent, we reinterpret them in terms of systematics templates. To isolate and track the systematics induced by the departure of the true beam from an ideal circular Gaussian profile, as done in \cite{Morshed_2026}, we can partition the timestream of Eq.~\eqref{eq:ideal_tod} as follows
\begin{equation}
        d(\Omega_t, \zeta_t, \psi) = \left[ {T} + {}_{0}^{T}Z + {}_{0}^{P}Z \right] + \left[ \left( \frac{{Q} + i{U}}{2} \right) e^{-i 2\psi}+ {}_{2}^{P}Z  + {}_{2}^{T}Z \right] e^{-i 2\zeta_t} + \left[ \left( \frac{{Q} - i{U}}{2}\right) e^{i 2\psi} + {}_{-2}^{P}Z  + {}_{-2}^{T}Z \right] e^{i 2\zeta_t} + \sum_{n \in \mathbb{Z} \setminus \{-2,0,2\}} \left[ {}_{n}^{T}Z + {}_{n}^{P}Z  \right] e^{-i n \zeta_t},
\end{equation}
where ${}_{n}^{T}Z$ and ${}_{n}^{P}Z$ with $n\in\mathbb{Z}$ represent the underlying spin leakage maps, respectively from temperature and polarization, defined as:
\begin{align}
    {}_{n}^{T}Z(\Omega) &= \sum_{\ell=0}^{\infty} \sum_{m=-\ell}^{\ell} (-1)^n \sqrt{\frac{4\pi}{2\ell +1}} {}_{0}a_{\ell m} \, {}_{0}b_{\ell, -n}^{\text{syste}\,\dagger} \, {}_{n}Y_{\ell m}(\Omega), \\
    {}_{n}^{P}Z(\Omega) &= \sum_{\ell=0}^{\infty} \sum_{m=-\ell}^{\ell} (-1)^n \sqrt{\frac{\pi}{2\ell +1}} \left( {}_{2}a_{\ell m} \, {}_{2}b_{\ell, -n}^{\text{syste}\,\dagger} + {}_{-2}a_{\ell m} \, {_{-2}}b_{\ell, -n}^{\text{syste}\,\dagger} \right) \, {}_{n}Y_{\ell m}(\Omega).
\end{align}
Here, $b_{\ell m}^{\text{syste}} \coloneqq b_{\ell m} - b_{\ell m}^{\text{cg}}$ quantifies the perturbation of the true beam from the reference circular Gaussian profile. Consistent with Eq.~\eqref{eq:beam-convolved-tod}, the quantities ${T}$, ${Q}$, and ${U}$ denote the true sky Stokes parameters smoothed by the reference circular Gaussian beam:
\begin{equation}
{T}(\Omega) = \sum_{\ell=0}^{\infty} \sum_{m=-\ell}^{\ell} {}_{0}a_{\ell m} \, {}_{0}b_{\ell 0}^{\text{cg}} \, {}_{0}Y_{\ell m}(\Omega), \qquad\qquad {Q}(\Omega) \pm i{U}(\Omega) = \sum_{\ell=0}^{\infty} \sum_{m=-\ell}^{\ell} {}_{\pm 2}a_{\ell m} \ {}_{\pm 2}b_{\ell, \mp 2}^{\text{cg}} \, _{\pm 2}Y_{\ell m}(\Omega).
\end{equation}
By comparing this expansion to Eq.~\eqref{eq:spin-timestream}, we can identify the effective spin maps required for the mapmaking equation, Eq.~\eqref{eq:mapmaking}:
\begin{align}
    \tilde{S}_{0} &= {T} + {}_{0}^{T}Z + {}_{0}^{P}Z, \\
    \tilde{S}_{2} &= \left( \frac{{Q} + i{U}}{2}  \right) e^{-i 2\psi} + {}_{2}^{P}Z +\prescript{T}{2}{Z}, \\
    \tilde{S}_{-2} &= \left( \frac{{Q} - i{U}}{2} \right) e^{i 2\psi}  + \prescript{P}{-2}{Z} + \prescript{T}{-2}{Z}, \\
    \tilde{S}_{n} &= {}_{n}^{T}Z + {}_{n}^{P}Z e^{-i2\psi}, \quad \text{for } n \in \mathbb{Z} \setminus \{0, 2,-2\}.
\end{align}
This decomposition shows that the departure from a circular Gaussian beam induces systematics terms at all spins. At $n=0,\pm2$, these terms add to the input temperature and polarization, thus they couple to $h_0$ in the discrete convolution of Eq.~\eqref{eq:mapmaking}, so they cannot be reduced by a more uniform scanning, which attenuates the higher-order $h$-maps. For $n \in \mathbb{Z} \setminus \{0, \pm2\}$, the departure instead introduces new spin terms that are absent for an ideal circular beam\footnote{This formulation only serves as an interpretation of the spin maps; in practice we compute them using Eq.~\eqref{eq:expr_spin_maps}.}.

Assuming no subpixel pointing offset between paired detectors (so that detectors in a pair share the same $h$-maps), this formulation recovers the known result (shown for instance in \cite{Mirmelstein:2020pfk}) that T-to-P leakage terms cancel out when considering pairs of orthogonal detectors that share the same beam. Since the detectors in a pair share the same ${}_{n}^{T}Z$ and $h$-maps, these terms cancel pair by pair in the mapmaking through the $e^{2 i \psi}$ factor in the polarization input spin terms, seen directly in Eq.~\eqref{eq:mapmaking}. Likewise, the temperature/polarization mixing terms of the mapmaking matrix, $\tilde{h}^{\text{tot}}_{\pm 2}$, vanish pair by pair thanks to the same $e^{2 i \psi}$ factor.

\section{Cross-polar response in beam harmonic decomposition}
\label{sec:crosspolar}

In Section~\ref{sec:formalism}, the beam harmonic expansion was presented assuming an ideal copolar detector response, i.e. the detector is exclusively sensitive to the electric field component aligned with its nominal polarization direction ($\tilde{\mathcal{E}}_{\mathrm{co}}$) and completely blind to the orthogonal cross-polar component ($\tilde{\mathcal{E}}_{\mathrm{cross}}$). In realistic optical systems, feedhorns and optical elements induce non-zero cross-polar leakage, modifying the effective real-space response of the detector.

Following \cite{Challinor:2000df}, the detector's real-space response can be decomposed into copolar and cross-polar field components:
\begin{align}
    \tilde{T}(\mathbf{r}) &= \langle |\tilde{\mathcal{E}}_{\mathrm{co}}(\mathbf{r})|^{2} + |\tilde{\mathcal{E}}_{\mathrm{cross}}(\mathbf{r})|^{2} \rangle, \\
    \tilde{Q}(\mathbf{r},\zeta + \psi) &= -\langle |\tilde{\mathcal{E}}_{\mathrm{co}}(\mathbf{r})|^{2} - |\tilde{\mathcal{E}}_{\mathrm{cross}}(\mathbf{r})|^{2} \rangle \cos 2 (\zeta + \psi) + 2\Re\langle \tilde{\mathcal{E}}_{\mathrm{co}}(\mathbf{r}) \tilde{\mathcal{E}}^{*}_{\mathrm{cross}}(\mathbf{r}) \rangle \sin 2 (\zeta + \psi), \\
    \tilde{U}(\mathbf{r},\zeta + \psi) &=  - \langle |\tilde{\mathcal{E}}_{\mathrm{co}}(\mathbf{r})|^{2} - |\tilde{\mathcal{E}}_{\mathrm{cross}}(\mathbf{r})|^{2} \rangle \sin 2 (\zeta + \psi) - 2\Re\langle \tilde{\mathcal{E}}_{\mathrm{co}}(\mathbf{r}) \tilde{\mathcal{E}}^{*}_{\mathrm{cross}}(\mathbf{r}) \rangle \cos 2 (\zeta + \psi),
\end{align}

where $\mathbf{r}$ denotes the position in the focal plane relative to the beam center, $\psi$ is the nominal detector orientation angle, and $\zeta$ is the local orientation angle. The brackets emphasize that we consider the time-averaged detector response.

The spin-weighted beam harmonic coefficients ${}_{\pm 2} b_{\ell m}$ are defined via the projection:
\begin{align}
    {}_{\pm 2} b_{\ell m} &= \int d\mathbf{r} \, \left( \widetilde{Q}(\mathbf{r},\psi) \pm i\, \widetilde{U}(\mathbf{r},\psi) \right) \, {}_{\pm 2}Y_{\ell m}^{\ast}(\mathbf{r}) \\
    {}_{0} b_{\ell m} &= \int d\mathbf{r} \, \widetilde{T}(\mathbf{r}) \, {}_{0}Y_{\ell m}^{\ast}(\mathbf{r}).
    \label{eq:b2_projection}
\end{align}
Substituting the expressions for $\widetilde{Q}(\mathbf{r})$ and $\widetilde{U}(\mathbf{r})$ into Eq.~\eqref{eq:b2_projection} yields the full expression of the beam harmonic coefficients in presence of both co-polar and cross-polar response:
\begin{align}
    {}_{\pm 2} b_{\ell m} &= -\int d\mathbf{r} \, 
    \left[ 
    \langle |\tilde{\mathcal{E}}_{\mathrm{co}}|^{2} - |\tilde{\mathcal{E}}_{\mathrm{cross}}|^{2} \rangle 
    \pm i\, 2\Re\langle \tilde{\mathcal{E}}_{\mathrm{co}} \tilde{\mathcal{E}}^{*}_{\mathrm{cross}} \rangle 
    \right] 
    e^{\pm i 2  \psi} \, {}_{\pm 2}Y_{\ell m}^{\ast}(\mathbf{r}) \\
    {}_{0} b_{\ell m} &= \int d\mathbf{r} \, \langle |\tilde{\mathcal{E}}_{\mathrm{co}}(\mathbf{r})|^{2} + |\tilde{\mathcal{E}}_{\mathrm{cross}}(\mathbf{r})|^{2} \rangle\, {}_{0}Y_{\ell m}^{\ast}(\mathbf{r}).
\end{align}

From these expressions, it is straightforward to see that in our $h$-map formalism, cross-polar effects alter only the numerical values of the input beam harmonic coefficients $b_{\ell m}$. Because these corrections are pre-absorbed into the effective spin sky maps $\tilde{S}_n(\Omega)$ via Eq.~\eqref{eq:expr_spin_maps}, accounting for realistic cross-polar beam leakage has no computational overhead during mapmaking or map-based beam convolution simulations.

\section{Comparison with perturbative templates of beam ellipticity}
\label{sec:pert_ell}
Here, we compare the results obtained with our exact beam harmonic approach to those from a perturbative method based on a Taylor expansion of the beam around the circular Gaussian case. In this framework, one can derive spin-map systematics templates of beam ellipticity from the coefficients of the second-order Taylor expansion of an elliptical beam \citep{Fabbian:2021hlw,Morshed_2026}, yielding a spin-2 systematics template. Using these perturbative elliptical beam templates in \smarties, with the same detector setup as in Section~\ref{sec:validation}, we compare the resulting maps to those from \lbs $4\pi$ convolution. Since the specific templates we used only model T-to-P leakage and not P-to-P effects, we adopt an unpolarized input sky, hence the polarized output consists solely of systematics.
\\
\begin{figure}[!h]
    \centering
    \includegraphics[width=0.9\linewidth]{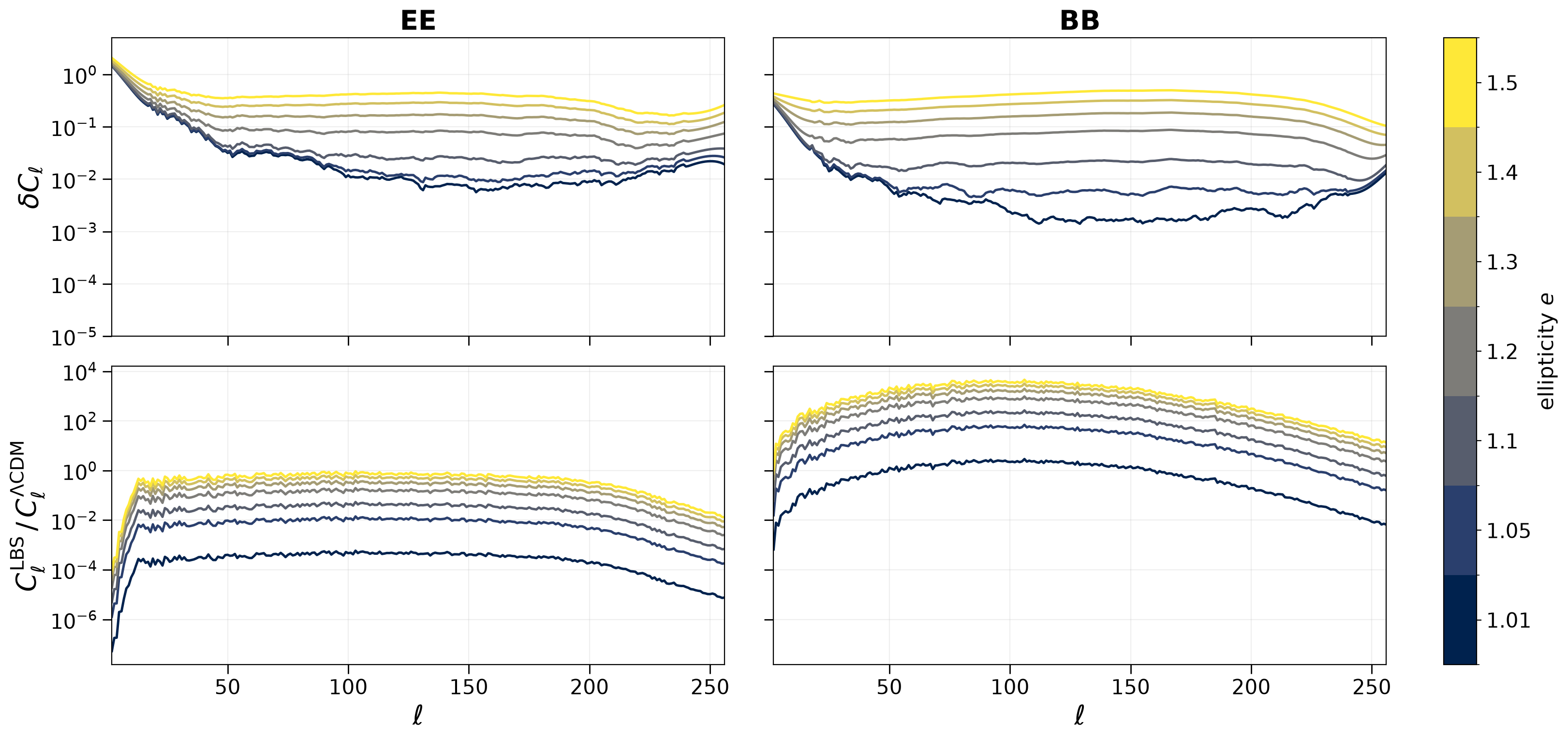}
    \caption{Top row: relative error $\delta C_\ell \coloneqq \left| C_\ell^{\rm LBS} - C_\ell^{\rm SMARTIES} \right| / C_\ell^{\rm LBS}$ on the systematics power spectra between \lbs $4\pi$ convolved maps and \smarties perturbative templates, both computed with an unpolarized input. Results are shown for several values of the beam ellipticity $e$, color-coded. Bottom row: \lbs systematics auto-spectra normalized by the theoretical $\Lambda$CDM $C_\ell$. A Savitzky-Golay filter with a $\ell=10$ window was applied to improve the readability of the low-ellipticity curves.}
\label{fig:pertubative_ell_beam}
\end{figure}
Figure~\ref{fig:pertubative_ell_beam} shows, in the top row, the relative error $\delta C_\ell \coloneqq \left| C_\ell^{\rm SMARTIES} - C_\ell^{\rm LBS} \right| / C_\ell^{\rm LBS}$ on the EE and BB systematics spectra, for several values of the beam ellipticity $e$, color-coded. The bottom row displays the corresponding \lbs systematics auto-spectra normalized by the theoretical $C_\ell$ of a \planck 2018 cosmology \citep{planck2016-l06}. The perturbative templates show good agreement with \lbs at the relative error level for small yet realistic ellipticities; $e=1.01$ corresponds to the smallest ellipticities estimated from ray-tracing simulations in the Simons Observatory LAT focal plane \citep{Gudmundsson:2020wja}. The situation degrades significantly for higher ellipticities: above $e=1.3$, the relative error exceeds 10\%, in a regime where the systematics amplitude is larger than that of the theoretical BB spectrum. This makes these templates unsuitable for simulating the systematics produced by significantly elliptical beams that could be featured in compact satellite experiments, such as those seen in \planck \citep{planck2013-p03c,planck2014-a05}.
\\
The performance of this perturbative approach could likely be improved by extending it to higher-order terms in the Taylor expansion, and thus higher spin orders. However, because derivatives of the beam translate to derivatives of the temperature field in the systematics templates, this would require computing higher-order derivatives of the temperature field, which can prove numerically impractical. Even with higher-order terms, the perturbative approach remains less flexible than directly using the beam harmonic coefficients, where it is straightforward to account for generic distortions of the beam shape. This further highlights the relevance of the exact formalism developed in this work as a robust and computationally cheap alternative to $4\pi$ TOD-based convolutions.

\section{Assessing the choice of \texorpdfstring{$m_{max}$}{m\_max}}
\label{sec:mmax_discussion}
\begin{figure}[h]
    \centering
    \includegraphics[width=1\linewidth]{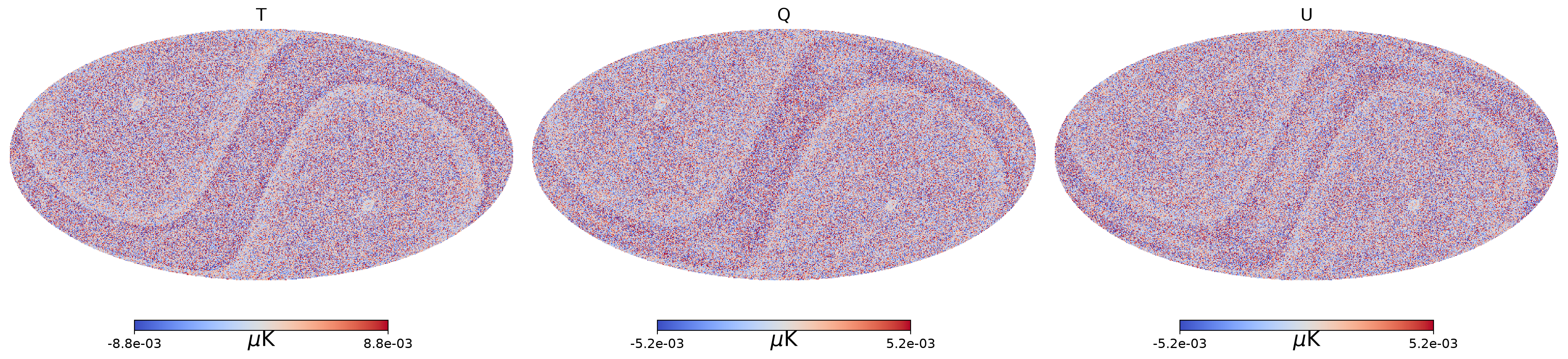}
    \caption{Difference between T,Q,U maps convolved with 100 GHz HFI beams from \lbs $4\pi$ convolution and \smarties map-based convolution. The reference \lbs output was produced with $m_{max}^\mathrm{LBS}=15$ while the \smarties one with $m_{max}^\mathrm{SMARTIES}=6$. The scanning pattern is clearly visible as a consequence of the omission of spin maps with spin $m_{max}^\mathrm{SMARTIES}<m\leq m_{max}^\mathrm{LBS}$. As quantified in Fig.~\ref{fig:mmax_truncation_test_hfi}, these residuals correspond, for the case plotted here, to a sub-percent mismatch in the systematics maps in terms of the cross-correlation factor.}
    \label{fig:residual_maps_mmax}
\end{figure}

During the validation, in the elliptical and HFI beams case, we truncate the $b_{\ell m}$ at $m_{max}=4$ or $6$, to keep the number of stored $h$-maps small. This modifies the geometry of the beam and introduces a mismatch with respect to a convolution computed with the full set of $b_{\ell m}$ (i.e., $m_{max}=\ell_{max}$). Although our formalism supports arbitrary $m_{max}$, it is important to determine the value required for sufficient precision, as the computational and storage costs grow with $m_{max}$. To quantify it, we consider a baseline of $4\pi$ convolved maps produced with a high $m_{max}^\mathrm{LBS}= 15$\footnote{The amplitude of the $b_{\ell m}$ drops fast enough with $m$ so that the mismatch between $4 \leq m_{max} \leq 8$ and $m_{max}=\ell_{max}$ is effectively the same as between $m_{max}=4,6$ and $m_{max}=15$.}, to maps produced with our map-based convolution using a lower $m_{max}^\mathrm{SMARTIES}=4$ or $6$. We are effectively omitting spin maps with a spin greater than $m_{max}^\mathrm{SMARTIES}+2$ in Eq.~\eqref{eq:mapmaking}. The omitted spin-maps coupling with the $h$-maps leaves a residual that traces the scanning pattern, as seen in Fig.~\ref{fig:residual_maps_mmax} for the HFI beams; the same patterns are seen when considering elliptical beams.

\begin{figure}[!ht]
    \centering
    \includegraphics[width=\linewidth]{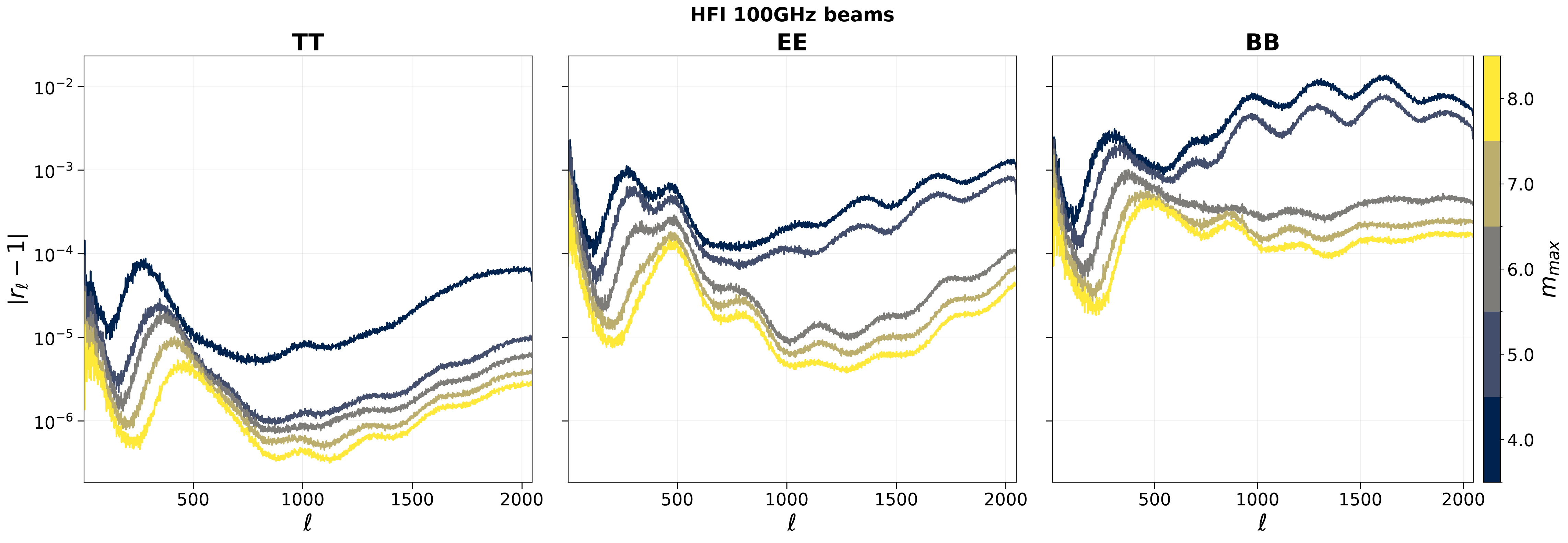}
    \caption{Deviation of the cross-correlation factor $r_\ell^{X(m_1) X(m_2)}$ from unity (Eq.~\eqref{eq:cross-corr}) between the systematics maps obtained with $m_{max}^\mathrm{SMARTIES}$ and with the baseline $m_{max}^\mathrm{LBS}=15$, for \planck HFI 100 GHz scanning beams. Results are shown for $4 \leq m_{max} \leq 8$.}
\label{fig:mmax_truncation_test_hfi}
\end{figure}

\begin{figure}[!ht]
    \centering
    \includegraphics[width=\linewidth]{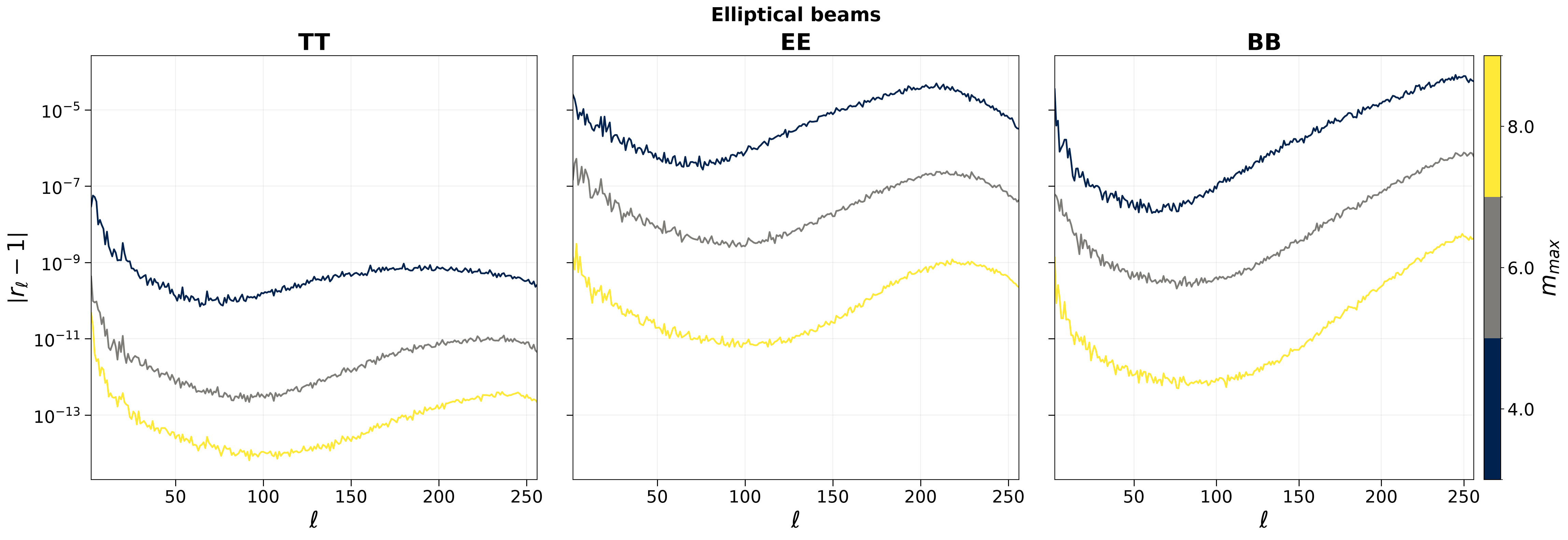}
 \caption{Deviation of the cross-correlation factor $r_\ell^{X(m_1) X(m_2)}$ from unity (Eq.~\eqref{eq:cross-corr}) between the systematics maps obtained with $m_{max}^\mathrm{SMARTIES}$ and with the baseline $m_{max}^\mathrm{LBS}=15$, for an elliptical beam with a 100 arcmin FWHM and ellipticity $e=1.3$. Results are shown for $ m_{max} \in \{4,6,8\}$.}
    \label{fig:mmax_truncation_test_ell}
\end{figure}

To quantify how much we lose in terms of simulating the systematics effects stemming from beam convolution, we decompose the observed maps into the input smoothed by the beam FWHM plus systematics. We then compute the cross-correlation factor between the systematics maps obtained with $m_{max}^\mathrm{SMARTIES}$ and with the baseline $m_{max}^\mathrm{LBS}=15$:
\begin{equation}
r_\ell^{X(m_1) X(m_2)} = \frac{C_\ell^{X(m_1) X(m_2)}}{\sqrt{C_\ell^{X(m_1) X(m_1)} \, C_\ell^{X(m_2) X(m_2)}}},
\label{eq:cross-corr}
\end{equation}
where $X \in \{T, E, B\}$, $m_1 = m_{max}^\mathrm{SMARTIES}$, and $m_2 = m_{max}^\mathrm{LBS} = 15$. The closer this factor is to unity, the more compatible the \smarties and \lbs systematics maps are. Figures~\ref{fig:mmax_truncation_test_hfi} and~\ref{fig:mmax_truncation_test_ell} display the cross-correlation factor deviation from unity for the \planck 100 GHz HFI beams and for elliptical beams with ellipticity $e=1.3$, respectively.

For the HFI beams (Figure~\ref{fig:mmax_truncation_test_hfi}), $|r_\ell - 1|$ exceeds $10^{-2}$ only in BB at $m_{max}^\mathrm{SMARTIES}=4$, indicating that the systematics maps are already recovered to a few-percent accuracy at this truncation; it drops by another two orders of magnitude when going to $m_{max}^\mathrm{SMARTIES}=8$. For elliptical beams with ellipticity $e=1.3$ (Figure~\ref{fig:mmax_truncation_test_ell}), at the higher end of what was measured for \planck LFI beams \citep{planck2014-a05}, the agreement is even better, with $|r_\ell - 1| \lesssim 10^{-3}$ at $m_{max}^\mathrm{SMARTIES}=4$. In both cases, the cross-correlation factor is already close to unity at the smallest truncation considered, confirming that a modest number of $h$-maps is sufficient to accurately reproduce the systematics when considering convolution by the main beam.

Overall, we obtain very good agreement at the systematics simulation level with small values of $m_{max}^\mathrm{SMARTIES}$ when considering the main beam convolution. In practice, the optimal value of $m_{max}^\mathrm{SMARTIES}$ will depend on the specific beam and the target accuracy.

\section{Leakage between scales}
\label{sec:scale_leakage}
\begin{figure}[!ht]
    \centering
    \includegraphics[width=\linewidth]{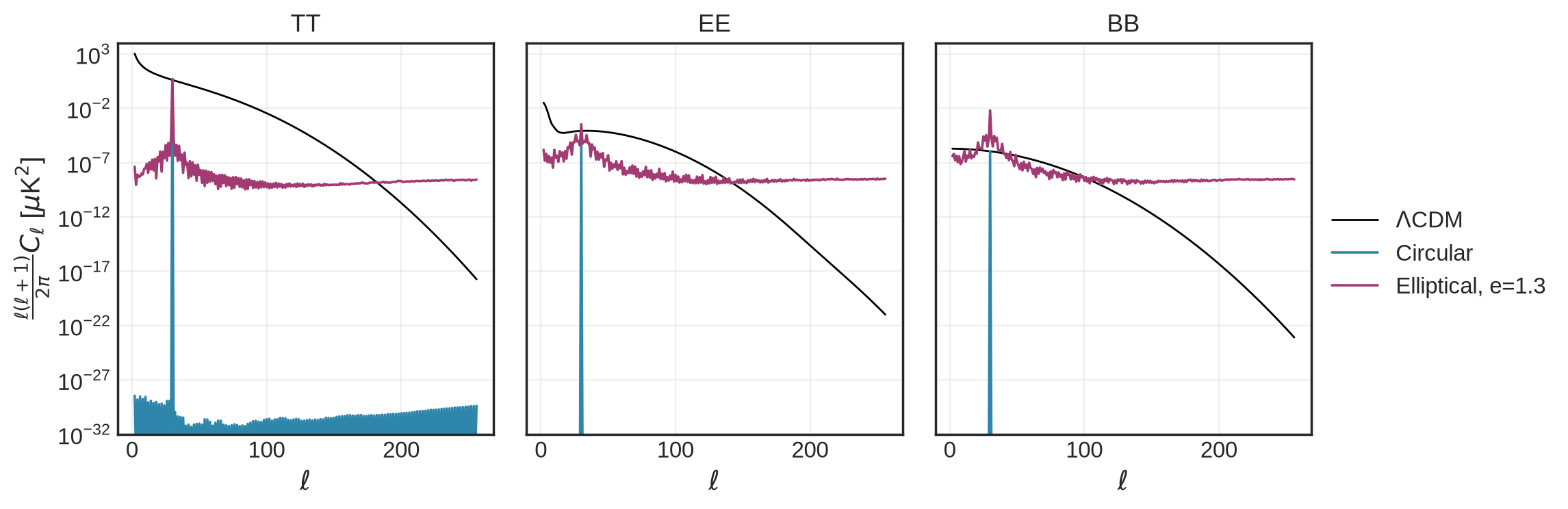}
    \caption{Power spectra of beam-convolved maps produced by \smarties with the input $C_\ell$ set to zero except at $\ell=30$. The full input spectrum from which $C_{30}$ is sampled, damped by a 200 arcmin Gaussian window function, is shown as a baseline.}
    \label{fig:leakage_scale}
\end{figure}

To illustrate the scale-mixing induced by non-circular beams, we perform a simple test using the same setup as in Section~\ref{sec:validation}, but with all $C_\ell$ of the \planck 2018 cosmology \citep{planck2016-l06} set to zero, except for $\ell=30$. Figure~\ref{fig:leakage_scale} shows the power spectra of the convolved maps produced by \smarties for an elliptical beam (30\% ellipticity) and a circular Gaussian beam, both with FWHM = 200 arcmin. For comparison, we also show the full input $C_\ell$ damped by a 200 arcmin Gaussian window function, from which we extracted the input $C_{30}$. While the circular beam produces a power spectrum with power only at $\ell=30$, as expected, the elliptical beam introduces significant power at all multipoles, clearly demonstrating scale mixing\footnote{We also ran \lbs $4\pi$ convolution for the corresponding cases and checked that it yields exactly the same results as \smarties.}.

At high $\ell$, the leakage from $\ell=30$ surpasses the damped input $C_\ell$, supporting the claim made in Section~\ref{sec:planck}, that the discrepancy between \quickpol and \smarties at high $\ell$ arises from the \quickpol assumption of no scale mixing.

We can explain this effect by considering only the temperature for simplicity\footnote{A similar, slightly more cumbersome, derivation can be worked out for the polarization signal.}. Non-circular beams result in observed temperature maps that contain non-zero spin maps due to the discrete convolution with $h$-maps in Eq.~\eqref{eq:mapmaking}. As a result, the observed temperature is constructed from spin-weighted harmonics with $s \neq 0$. The observed temperature harmonic coefficients are defined as

\begin{eqnarray}
    {_{\hat{T}} a_{\ell m}} &=& \int_{4\pi} d\Omega \ \hat{T}(\Omega) \ {}_0Y_{\ell m}^{\dagger}(\Omega) \\
    &=& \int_{4\pi} d\Omega \left[ \sum_{j=1}^{N_{\text{det}}} \frac{N^j_{\text{hits}}}{N^{\text{tot}}_{\text{hits}}} \sum_{k' = - \infty}^{\infty} \tilde{h}_{- k'}^j \tilde{S}_{k'}^j \right] {}_0Y_{\ell m}^{\dagger}(\Omega) \\
    &=& \int_{4\pi} d\Omega \sum_{j=1}^{N_{\text{det}}} \frac{N^j_{\text{hits}}}{N^{\text{tot}}_{\text{hits}}} \sum_{k' = - \infty}^{\infty} \tilde{h}_{- k'}^j \sum_{\ell '=|k'|}^{\infty} \sum_{m'=-\ell'}^{\ell'} \\ && (-1)^{k'} \sqrt{\frac{4\pi}{2\ell' +1}} \left( {}_{0}a_{\ell' m'} \, {}_{0}b_{\ell', -k'}^{\dagger} +\frac{1}{2}  {}_{2}a_{\ell' m'} \,{}_{2}b_{\ell' ,-k'}^\dagger +\frac{1}{2} {}_{-2}a_{\ell' m'} \,{}_{-2}b_{\ell', -k'}^\dagger \right)\ \ {}_{k'}Y_{\ell', m'}(\Omega)  {}_0Y_{\ell m}^{\dagger}(\Omega)
\end{eqnarray}
where, to keep the equation tidy, we assumed $\tilde{h}^{\text{tot}}_{\pm2}=0$ in Eq.~\eqref{eq:mapmaking}, so that the temperature row of the mapmaking system decouples, and used $\tilde{h}^{\text{tot}}_{0}=1$.
Therefore, $_{\hat{T} } a_{\ell m}$ a priori depends on the input $a_{\ell' m'}$ with $(\ell', m') \neq (\ell, m)$. This leads to scale mixing when computing the power spectra.

\section{\sforp: A tool to produce beam-convolved simulated \planck maps}
\label{sec:planck_code}

To facilitate the generation of realistic, beam-convolved synthetic \planck maps, we developed \sforp, a Python package built as a high-level wrapper around \smarties. The pipeline requires three primary inputs:
\begin{enumerate}
    \item The \quickpol pointing moments (polmoments),
    \item The detector beam spherical harmonic coefficients ($b_{\ell m}$),
    \item The spherical harmonic coefficients ($a_{\ell m}$) of the input sky.
\end{enumerate}

The input polmoments, originally stored in detector-centric \textit{Pxx} coordinates, are automatically transformed into the $h$-map conventions required by \smarties and weighted according to the relative \planck detector weights. If the supplied beam files do not contain explicit polarization information, \sforp generates the polarized components under the copolar approximation (${}_{\pm 2}b_{\ell m} = {}_{0}b_{\ell ,m \pm 2} \ e^{\pm 2i\psi}$).

Once the $h$-maps and effective beam harmonics are prepared, \sforp invokes \smarties to evaluate the corresponding spin sky maps and perform the mapmaking inversion (Eq.~\eqref{eq:mapmaking}). The output consists of full-sky, beam-convolved temperature and polarization ($T, Q, U$) maps in \healpix format. 

The required auxiliary \planck data (polmoments and $b_{\ell m}$ files) are accessible via NERSC\footnote{Located at \texttt{/global/cfs/cdirs/cmb/data/planck2020/npipe/aux} for the NPIPE release. The polmoments are in \texttt{aux/polmoments}, the beams are in \texttt{aux/beams}.}, and an example notebook demonstrating the complete workflow is available in the repository.
 \end{appendix}
\end{document}